\documentclass[A4paper,12pt]{article}
\usepackage[T1]{fontenc}
\usepackage[english]{babel}
\usepackage[cp1250]{inputenc}
\usepackage{epsfig}
\usepackage{amsfonts}
\usepackage{fancyhdr}
\usepackage{syntonly}
\usepackage{graphicx}

\begin{document}
\selectlanguage{english}

\begin{titlepage}
\begin{center}
\vspace*{3cm}

\begin{title}
\bold {\Huge On the multiplicity distributions\\
\vspace{0.3cm} at LHC energies
 }
\end{title}

\vspace{2cm}

\begin{author}
\Large K. FIA{\L}KOWSKI\footnote{e-mail address:
fialkowski@th.if.uj.edu.pl}, R. WIT\footnote{e-mail address:
romuald.wit@uj.edu.pl}

\end{author}

\vspace{1cm}

{\sl M. Smoluchowski Institute of Physics\\ Jagellonian University \\

30-059 Krak{\'o}w, ul.Reymonta 4, Poland}

\vspace{2cm}

\begin{abstract}
The ALICE and CMS data on the multiplicity distributions are
compared with the lower energy data and with the results from the
8.142 version of the PYTHIA MC event generator with two tunings. The
ALICE data for moments are used to calculate the factorial
cumulants. It is suggested that the data on moments or cumulants are
well suited to specify the optimal tuning of the model parameters.
\end{abstract}

\end{center}

\vspace{2cm}

PACS:   13.85.-t, 13.90.+i \\

{\sl Keywords:}  LHC, multiplicity distributions  \\

\end{titlepage}

\section{Introduction}

With the advent of LHC data it became possible to investigate the
multiplicity distributions at the CM energies beyond $2$ TeV. It is
very interesting to check how well the default versions of MC
generators describe the minimum bias events in this energy range. In
particular, it is tempting to look for a best tuning of the model
parameters using only the data from the multiplicity distributions.
\par
In a recent note \cite{KFRW} we have discussed the energy dependence
of the central density (defined by the average charged multiplicity
in a central bin in pseudorapidity). We have shown that, contrary to
some claims, the fast increase with energy observed in the ALICE
data \cite{ALICE2} is not unexpected. In fact, we found that the
default version of PYTHIA 8.135 generator \cite{SMS}, \cite{SMS2}
predicts a too fast increase, but with some tuning the data may be
well described. The moments for three selected pseudorapidity bins
were also compared with data \cite{ALICE}. The qualitative agreement
was observed and the spread of results for two different tunings was
surprisingly small.
\par
In this paper we discuss the data on the multiplicity distributions
listed above as well as the data from the CMS experiment \cite{CMS}
compared with some lower energy data \cite {UA5}, \cite{UA1},
\cite{NA22}. We use here the new version of PYTHIA 8.142 \cite{TS}
with the default tuning and with the tuning used by default in the
older version 8.135. Let us note that the PYTHIA 8.142 version is
significantly changed compared to its earlier versions (PYTHIA 8.107
and 8.135) which we have used in a previous publication \cite{KFRW}.
The changes, concerning mainly the final state radiation, are
motivated by the discrepancy between the Tevatron data for
"underlying event" and the model results. The qualitative
conclusions of our note \cite{KFRW} are, however, unchanged.
\par
In the next section, we compile for convenience the formulae
defining the moments. factorial moments and factorial cumulants of
the multiplicity distributions. Then we recall the results from Ref.
\cite{KFRW} and compare them with the factorial cumulants calculated
from the data and MC generators. We will see that that the scaled
factorial cumulants $K_q$, calculated from the published values of
the standard scaled moments $c_q$ and the average multiplicities
$\overline{n}$, exhibit much smaller spread than suggested by the
published values of the uncertainties of $c_q$ and $\overline{n}$.
Thus it would be more reasonable to use the scaled factorial
cumulants of the multiplicity distributions to test the specific
models of the high energy collisions. Finally, we discuss the energy
dependence of the average multiplicity and second moment in a wider
energy range, as presented in the CMS paper \cite{CMS}. The last
section contains some conclusions and the outlook.

\section{Moments and cumulants}

The multiplicity distributions are often parametrized in terms of
moments. This facilitates the comparison with models and allows for
a simple description of the energy dependence. The crucial problem
is the proper choice of the set of moments to be used. A standard
first choice is to use simple power moments defined by
$$\overline n^q = \Sigma n^q P(n)$$
or their scaled version
$$c_q = \frac {\overline{n^q}}{\overline{n}^q}.$$
\par
These moments are easy to calculate and (for moderate $q$) depend
quite uniformly on the probabilities, although obviously the lowest
multiplicities are suppressed, and the high multiplicity tail is
enhanced. The use of $\overline{n}$ and a few lowest $c_q$ moments allows
to parametrize the multiplicity distribution quite satisfactorily.
In the high energy (high $\overline{n}$) limit these moments allow to describe
the "KNO scaling function" $\Psi (x)$ \cite{KNO} defined by
$$\Psi (x) = \lim_{\overline{n} \rightarrow \infty} P(n)/\overline{n}$$
where $x=n/\overline{n}$. Obviously, in this limit
$$c_q = \int z^q \Psi (z) dz.$$
 However, for the distributions in small phase space bins
the power in the denominator of the formula for the scaled moments
results in the development of trivial singularities. Thus since some
time one uses more often the factorial moments. If we define the
factorial quotient
$$n_q = \frac{n!}{(n-q)!}$$
the corresponding standard and scaled factorial moments are,
respectively
$$\overline{n_q} = \Sigma n_q P(n)$$
and
$$F_q = \frac{\overline{n_q}}{\overline{n}^q}.$$
The factorial moments of the order $q$ are the integrals of the
q-particle densities for identical particles. Obviously,
$\overline{n_1}=\overline{n}$ is the integral of the single particle
density. For the smooth phase space distributions the scaled
factorial moments behave smoothly for the bin size decreasing to
zero, and the possible power increase is a signal for intermittency
\cite{BP}. However, the drawback of the definition of the higher
factorial moments is their independence on the lower end of the
multiplicity distribution. Moreover, the scaled factorial moments of
different order are strongly correlated.
\par
Therefore it is preferable to parametrize the multiplicity distributions
by the factorial cumulants. They are defined in a compact way by the
generating function
$$G(z) = \Sigma z^n P(n).$$
The factorial cumulants $f_q$ (called also "Mueller coefficients" \cite{AM}) are defined by
$$f_q = \frac{d^q (lnG(z))}{dz^q} \Big|_{z=1}$$
to be compared with an analogous definition of the factorial moments
$$\overline{n_q} = \frac{d^q G(z)}{dz^q} \Big|_{z=1}.$$
The scaled factorial moments are defined in a usual way
$$K_q = \frac{f_q}{\overline{n}^q}.$$
By definition, for $q=1$ all the scaled moments $c_q,~F_q$ and $K_q$ are equal one.
Another name for the factorial cumulants $f_q$ is "the correlation integrals",
as they are the integrals of the correlation functions of the order $q$.
Therefore they measure the genuine multiparticle correlations. For the
uncorrelated emission all the factorial cumulants for $q>1$ vanish. If
there are only two-particle correlations, $f_q=0$ for $q>2$. The values of
the factorial cumulants of different orders are uncorrelated.
\par
One may add that another set of moments was advocated \cite{ID}
$$H_q = K_q/F_q = f_q/\overline{n_q}.$$
These moments were shown to have strongly reduced statistical uncertainties
even for the order up to $q=10$. However, for $q<5$ they are not very practical
to use, as their values quickly decrease with increasing $q$.
\par
A practical difficulty in using the factorial cumulants is the complexity
of the formulae for their errors, or, more precisely, for their
statistical uncertainties. For the average multiplicity one uses
a simple estimate of the uncertainty
$$\Delta \overline{n} = D/\sqrt N$$
where $D$ is the dispersion, and $N$ is the total number of measured events.
Analogous simple formulae exist for the higher standard moments $\overline{n^q}$.
However, for the scaled moments, and especially for cumulants,
the corresponding formulae are more complicated. Moreover, usually they overestimate
significantly the observed spread of experimental results.
\par
There is a simple explanation of this fact. The formula for $\Delta \overline{n}$ was obtained
from a simple prescription for the statistical uncertainty of a parameter
$$\Delta A = \sqrt{\Sigma (\partial A/\partial N_n)^2 \cdot (\Delta N_n)^2}$$
with $\Delta N_n = \sqrt {N_n}$. This prescription results from the
assumption that the measured numbers of events with different
multiplicities $N_n$ are uncorrelated, and their errors are purely
statistical. These assumptions were reasonable e.g. for the hydrogen
bubble chamber experiments, where the full solid angle was available
for the measurements of tracks, and the multiplicity of charged
particles (always even for charged beam) was unambiguously measured.
This is certainly not the case for a colliding beam experiment with
electronic detectors, where the multiplicity distribution is
measured in the restricted bin of phase space. The uncertainty of a
measurement of the variables defining this bin as well as the
effects of the track splitting and joining due to the imperfection
of the detector result in non-statistical errors and in the strong
correlations between the numbers of events with different
multiplicities. Neither a simple formula for $\Delta \overline{n}$
presented above, nor the complicated formulae for the uncertainties
of the scaled factorial cumulants (derived from the same
prescription for the statistical uncertainties of the numbers of
events) are reliable.
\par
Therefore the realistic estimate of the uncertainties of the
parameters of the multiplicity distribution requires the full
knowledge of the detector and, in particular, the measurement of the
correlation matrix for the multiplicities. This can be done only by
the authors of the experiment. Readers cannot translate them
reliably into a different set of parameters, since their
uncertainties will be unknown.

\section {Moments and factorial cumulants}
In a recent paper \cite{ALICE} the ALICE collaboration has presented
the values of average multiplicities and scaled moments $c_q$ for
$q=2,3,4$ at two CM energies $0.9$ and $2.36$ TeV for three choices
of the central pseudorapidity bin widths: $\Delta \eta <1,~\Delta
\eta <2$ and $\Delta \eta <2.6$.
\par
In Table 1 and in Fig.1 we show the experimental values of the
$\overline{n}$ from the "non-single-diffractive" (NSD) ALICE data at
$900$ GeV and $2.36$ TeV and the corresponding values calculated
from the PYTHIA 8.142 default version. For each point we have
generated $10^5$ events. In all tables the numbers in parentheses
denote the statistical and systematic errors.

\begin{table}[!h]
\caption{Average multiplicities for three choices of rapidity bins
from ALICE and two versions of PYTHIA 8.142 at {\it 0.9} and {\it
2.36} TeV.} \label{Averages}
\begin{center}
\resizebox{0.85\textwidth}{!}{%
\begin{tabular}{||c|c|c|c|c|c|c||}
\hline
 \hline
        $ \eta $ range & ALICE $0.9$ TeV  &  P8.142&P8.142/135 &ALICE $2.36$ TeV  &  P8.142&P8.142/135 \\
\hline \hline
     $\mid \eta \mid <0.5$ &3.60(2)(11)&3.57&3.87 &4.47(3)(10)&4.45&4.75  \\
\hline
     $\mid \eta \mid <1.0$ &7.38(3)(17)&7.27 &7.88 &9.08(6)(29)&9.04 &9.66  \\
\hline
     $\mid \eta \mid <1.3$ &9.73(12)(19)&9.57 &10.35 &11.86(22)(45)&11.89 &12.68  \\
\hline
 \hline
\end{tabular}
}
\end{center}
\end{table}

\begin{figure}[h!]
\centerline{ \epsfig{figure=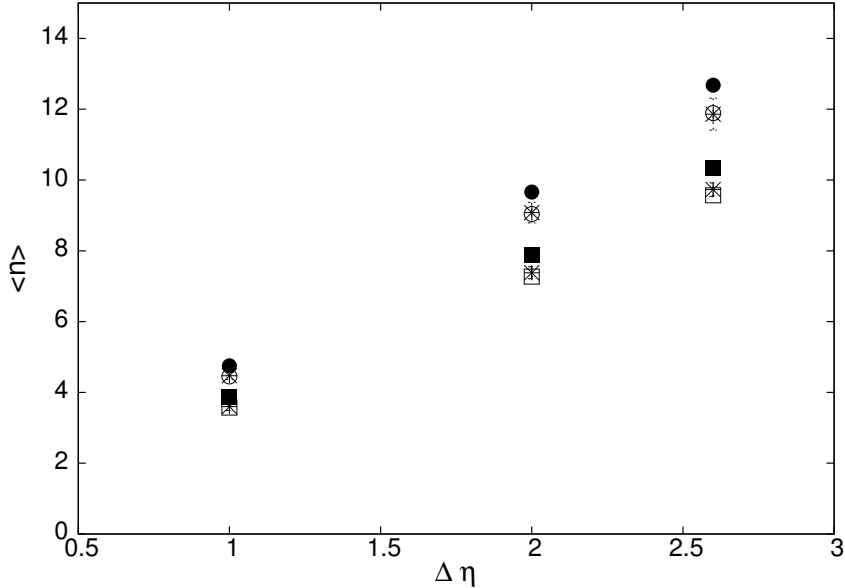,height=8.0cm}}
\caption{\footnotesize \label{Omega} The average multiplicity from
the ALICE data at {\it 0.9} and {\it 2.36} TeV (asterisks with error
bars), from PYTHIA 8.142 (open squares and circles) and from PYTHIA
8.142/135 (full squares and circles) as a function of the
pseudorapidity bin width}
\end{figure}

\par
The data for the average multiplicities agree well with the model
(remember that we are using the default version of PYTHIA 8.142
without any tuning). The agreement is significantly better than that
for the central densities of charged particles for inelastic events
with at least one particle in the central bin, measured by ALICE at
$0.9,~2.36$ and $7$ TeV \cite{ALICE2}, although in this case the
increase with energy is also reasonably well described, as seen in
Table 2 and in Fig.2.

 \begin{table}[h!]
\caption{Central density: data and the results for two versions of
PYTHIA 8.142} \label{PYTH1} \vspace{10pt}
\begin{center}
\resizebox{0.75\textwidth}{!}{%
\begin{tabular}{||c|c|c|c||}
 \hline
 \hline
       Energy (TeV)& ALICE &  PYTHIA 8.142 d &PYTHIA 8.142/135d \\
\hline \hline
          0.90 & 3.81(1)(7) & 3.58 & 3.86    \\
\hline
  2.36 & 4.70(1)(11) & 4.41 & 4.69 \\
\hline
     7.00 & 6.01(1)(20)  & 5.80 & 6.14 \\
\hline \hline

\end{tabular}
}
\end{center}
\end{table}

\begin{figure}[h!]
\centerline{ \epsfig{figure=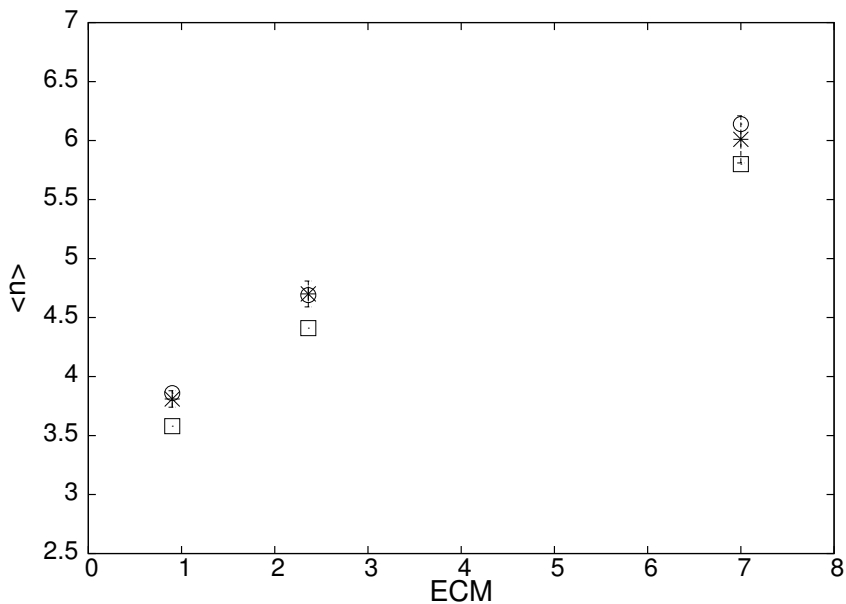,height=8.0cm}}
\caption{\footnotesize \label{INEL} Central density: data (asterisks
with error bars), PYTHIA 8.142 default (squares) and PYTHIA 8.142
with 8.135 tuning (circles)  as a function of the CM energy.}
\end{figure}

 \par
 We have repeated the same calculations for the PYTHIA 8.142 with
 a different set  of the model
 parameter values: the default values from the PYTHIA 8.135 version are taken.
Let us remind here that the tuned parameters refer to the formulae
used in the description of multiple scattering. The regularization
of the (divergent) QCD cross section is done by the introduction of
a factor
$$F(p_T)=\frac{p_T^4}{(p_{T0}^2+p_T^2)^2}$$
where
$$p_{T0}=pT0Ref\Big(\frac{ecmNow}{ecmRef}\Big)^{ecmPow}$$
and $ecmNow$ is the CM energy in GeV. The default values in PYTHIA
8.135 are $2.0$ for the $pT0Ref$, $1960.0$ for  $ecmRef$ and $0.16$
for $ecmPow$. In PYTHIA 8.142 the corresponding values are $2.15$,
$1800.0$ and $0.24$. Moreover, the default version of PYTHIA 8.142
uses a simplified profile of the parton density in the impact
parameter given by a Gaussian curve, whereas the standard earlier
versions were using two Gaussians (with two extra parameters for the
ratios of their slopes and weights). We have found that in this case
the results for the average multiplicities
 are reversed: there is a perfect agreement for central densities in
 the "INEL>0" sample (see Table 2) and a slight overestimation for the NSD sample.

\par
 The difference between Tables 1 and 2 shows that the ALICE procedures give for
 the PYTHIA events practically the same average multiplicity in the
 "NSD" and "INEL>0" samples, whereas experimentally the second sample
 has higher average multiplicity, which suggests lower contribution
 from the diffractive events. Remember that for PYTHIA we use the same definition of "NSD"
 and "INEL>0" events as in the data. This means that we generate all
 the classes of events (non-diffractive, single diffractive and
 double diffractive) and then remove the events which do not satisfy
 the conditions defined in the ALICE procedures.

\par
The formulae listed in the previous section allow to express the
scaled factorial cumulants in terms of the scaled moments and
average multiplicity. For the lowest values of $q$ we have
$$K_2=F_2-1,~F_2=c_2-1/\overline{n},~K_3=F_3-3F_2+2,~F_3=c_3-c_2/\overline{n}+2/{\overline{n}}^2$$
and
$$K_4=F_4-4F_3-3F_2^2+12F_2-6,~F_4=c_4-6c_3/\overline{n}+11c_2/{\overline{n}}^2-6/{\overline{n}}^3.$$
For the higher values of $q$ it is more practical to use a
recurrence formula expressing $K_q$ by $F_q$ and the $K$ moments of
the lower order
$$K_q=F_q - \sum_i \frac{(q-1)!}{(i-1)!(q-i)!} K_{q-i}F_i.$$
\par
In Table 3 we show the values of the $c_q$ and $K_q$ moments at
$900$ GeV, and in Table 4 the same results at $2.36$ TeV.

\begin{table}[!h]
\caption{Scaled moments and factorial cumulants for three choices of
rapidity bin from ALICE and two versions of PYTHIA 8.142 at {\it
0.9} TeV.} \label{Moments}
\begin{center}
\resizebox{0.9\textwidth}{!}{%
\begin{tabular}{||c|c|c|c|c|c|c|c|c||}
\hline
 \hline
       $ \eta $ range &$~~c_q~~$& ALICE  &  P8.142& P8.142/135    &$~~K_q~~$ &ALICE  &  P8.142& P8.142/135 \\
\hline \hline
     $\mid \eta \mid <0.5$ &$c_2$&1.96(1)(6) & 1.73 & 1.85          &  $K_2$ & 0.68 & 0.45 &  0.59        \\
\hline

    $\mid \eta \mid <1.0$ & $c_2$&1.77(1)(4) & 1.56&  1.70     &$K_2$ & 0.63 & 0.42   &   0.57   \\
\hline
  $\mid \eta \mid <1.3$ &$c_2 $ &  1.70(3)(7) & 1.51 & 1.65&    $K_2 $ &0.60 & 0.40   &  0.55        \\
\hline
     $\mid \eta \mid <0.5 $ &$c_3$ &5.35(6)(31)&  4.16& 4.93    & $K_3$ &0.82&  0.50    &  0.85   \\
\hline
 $\mid \eta \mid <1.0 $ &$c_3$ & 4.25(3)(20)&  3.29 &   4.11   &$K_3$ &0.66&  0.42   &   0.78       \\
\hline
  $\mid \eta \mid <1.3 $ &$c_3$ &3.91(10)(15)&  3.04 &  3.84   & $K_3$ &0.62&  0.38    &  0.73   \\
\hline
  $\mid \eta \mid <0.5$ &$c_4$&  18.3(4)(1.6)& 12.7 &   16.8   & $K_4$&1.13& 0.70   &  1.39    \\
\hline
   $\mid \eta \mid <1.0$ &$c_4$& 12.6(1)(9)& 8.65&   12.6   &$K_4$& 0.82& 0.50 &     1.24  \\
\hline
  $\mid \eta \mid <1.3$ &$c_4$&  10.9(4)(6)& 7.60&  11.3  & $K_4$& 0.57& 0.43     &  1.10    \\
 \hline
 \hline
\end{tabular}
}
\end{center}
\end{table}

\begin{table}[!h]
\caption{Scaled moments and factorial cumulants for three choices of
rapidity bin from ALICE and two versions of PYTHIA 8.142 at {\it
2.36} TeV.} \label{Moments2}
\begin{center}
\resizebox{0.9\textwidth}{!}{%
\begin{tabular}{||c|c|c|c|c|c|c|c|c||}
\hline
 \hline
    $ \eta $ range &$~~c_q~~$& ALICE  &  P8.142& P8.142/135    & $~~K_q~~$ &ALICE  &  P8.142& P8.142/135 \\
\hline \hline
   $\mid \eta \mid <0.5$ &$c_2 $ & 2.02(1)(4) & 1.75&  1.90    & $K_2 $ & 0.80 & 0.53   &   0.69    \\
\hline
   $\mid \eta \mid <1.0$ &$c_2 $ & 1.84(1)(6) & 1.61 & 1.76    & $K_2 $ & 0.73 & 0.50   &    0.65  \\
\hline
  $\mid \eta \mid <1.3$ & $c_2 $& 1.79(3)(7) & 1.56 &   1.71   & $K_2 $ & 0.71 & 0.48    &   0.64  \\
\hline
     $\mid \eta \mid <0.5 $ &$c_3$ &5.76(9)(26)&  4.31 & 5.25   &  $K_3$ & 1.12&  0.64    &  1.08     \\
\hline
 $\mid \eta \mid <1.0 $ &$c_3$ & 4.65(6)(30)&  3.56 &   4.46    &$K_3$ & 0.88&  0.55    &    0.98        \\
\hline
 $\mid \eta \mid <1.3 $ &$c_3$ & 4.35(16)(33)&  3.32 &  4.23    & $K_3$ & 0.79&  0.51   &    0.93      \\
\hline
  $\mid \eta \mid <0.5$ & $c_4$& 20.6(6)(1.4)& 13.4 &    18.6  &$K_4$& 1.77& 0.94    &    2.00     \\
\hline
  $\mid \eta \mid <1.0$ & $c_4$& 14.3(3)(1.4)& 9.80 &    14.4   &$K_4$& 0.98& 0.71   &     1.68      \\
\hline
  $\mid \eta \mid <1.3$ & $c_4$& 12.8(7)(1.5)& 8.75 &    13.1    &$K_4$&  0.83& 0.61   &    1.51       \\
 \hline
 \hline
\end{tabular}
}
\end{center}
\end{table}

These data show a few simple regularities:
\begin{enumerate}
\item
The values of the $c_q$ moments increase with the value of $q$ and
with energy, but decrease with the increasing pseudorapidity bin
width. The average multiplicity, as expected, increases with the bin
width and energy.
\item
The regularities listed above hold for $q=4$ even in the cases, when
the experimental errors given by the authors exceed the differences
between the data for different energies or different bin widths.
This is not so surprising for the bin width dependence, as the data
are here clearly quite strongly correlated. The presence of a
similar effect in the energy dependence seems to suggest that the
systematic errors at two energies are also correlated.
\item
 The values of $c_q$ in the two tunings differ by $0.15,~0.9$ and
 $5$ for $q=2,~3$ and $4$, respectively. In contrast, the values of
 $K_q$ differ much less for $q>2$. The model with the "wrong" tuning
 is compatible with data for $q>2$. This suggests that by a
 more refined tuning one may get a reasonable agreement with data
 not only for average multiplicities, but also for higher moments.
 Using the factorial cumulants we find a smaller spread
 of the values both in the model and in the data.
 \item
 The moments of the multiplicity
 distributions are systematically underestimated in the default tuning of
 PYTHIA 8.142. For
 the $c_q$ moments the difference between the data and the model
 values is around $0.2$, $1$ and $5$ for $q=2,~3$ and $4$,
 respectively, and increases weakly with the energy. For the $K_q$
 moments the trend is the same, but the differences for $q>2$ are much smaller:
 only in one case the difference is bigger than $0.5$.
 For the 8.135 tuning the situation is much more involved. All
 the values of the moments are now significantly higher. Whereas for
 $c_q$ they are still lower than in the data, the difference is
 really significant only for $q=2$. For the $K_q$ the values are
 below the data for $q=2$, and above the data for $q=4$.
\end{enumerate}
\par
It would be highly desirable to calculate reliably the experimental
errors of the scaled factorial cumulants to see how significant is
the difference between the model and data seen in Tables 3 and 4. We
have checked that the statistical uncertainties of the model results
for a given set of parameters are negligible: by increasing the
statistics by a factor of ten we do not change the values from the
Tables by more than a few percent. In all cases the observed
fluctuations for $c_q$ are negligible compared with the experimental
errors. However, as noted above, tuning the model parameters allows
to change the results sufficiently to hope for the agreement with
data.

\section {Energy dependence from SPS to LHC}

The CMS collaboration has also measured the multiplicity
distributions for non-single-diffractive (NSD) events at the CM
energies of $0.9$, $2.36$ and $7$ TeV \cite{CMS}. Contrary to the
most of published results, where the "NSD" events are defined just
by  giving the trigger conditions, the CMS data are extrapolated and
corrected to remove single diffraction. Since we are unable to
repeat this procedure in detail, the precise comparison with MC
results is beyond our ability. We have to rely on the effectiveness
of the CMS procedure and to compare their data with the events
generated in PYTHIA as non-single-diffractive.

\begin{figure}[h!]
\centerline{ \epsfig{figure=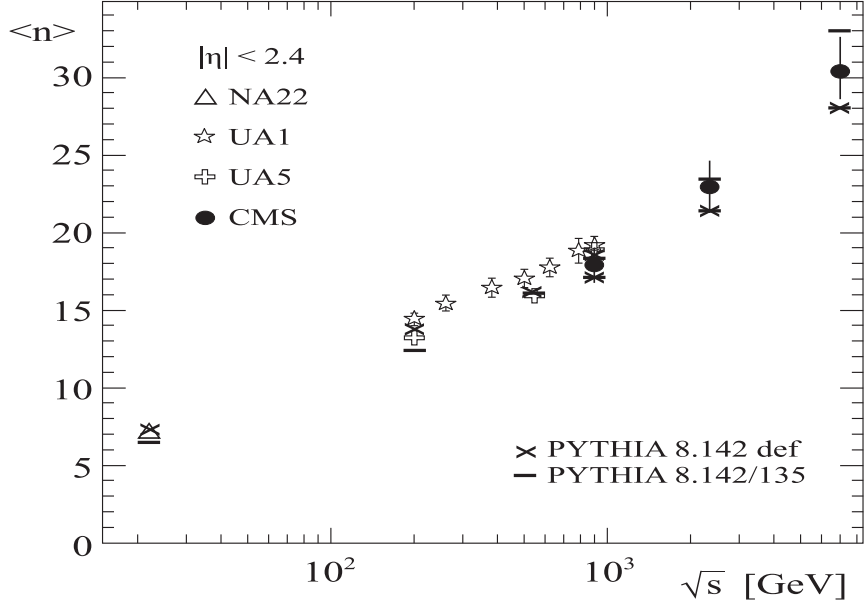,height=8.0cm}}
\caption{\footnotesize \label{NAV} The average multiplicity in the
$\mid \eta \mid < 2.4$ range from the NA22 data (triangle), UA5 data
(open stars), UA1 data (open crosses) and CMS data (black dots).
PYTHIA 8.142 predictions for NA22, UA5 and CMS with default and
8.135 tunings are shown as x-s and bars.}
\end{figure}

\par
The CMS paper refers also to the published data from the lower
energies: the CERN collider data from UA5 \cite{UA5} and UA1
\cite{UA1} collaborations and the SPS data from the EHS/NA22
collaboration \cite{NA22}. Thus it is possible to check if the
energy evolution in the wider range is reasonably described by the
PYTHIA 8 generator, and what are the differences between the
different tunings in this range. For the sake of transparency, the
UA1 data are shown only in the figure containing the average
multiplicities.

\par
In Fig.3 we show the data for the average multiplicity in the range
$\mid \eta \mid <2.4$ ($2.5$ for lower energies) and the PYTHIA
8.142 results with two tunings described above. At the CERN collider
energies the PYTHIA predictions are shown only with the UA5 trigger
conditions. We see that the MC results agree quite well with the
observed trend of the data and the values from the two tunings
bracket the experimental results.

\begin{figure}[h!]
\centerline{ \epsfig{figure=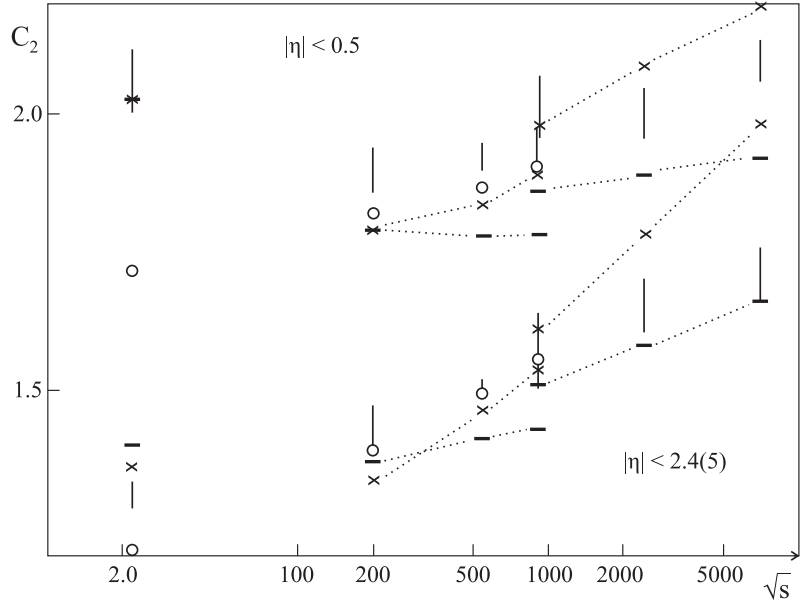,height=8.0cm}}
\caption{\footnotesize \label{C2} The $c_2$ moment in the $\mid \eta
\mid < 0.5$ and $\mid \eta \mid < 2.4$ range from the NA22, UA5 and
CMS data (vertical bars). PYTHIA 8.142 predictions with default and
8.135 tunings are shown as x-s and bars, respectively. The points
for UA5 and CMS data are connected by dotted lines to guide the
eye.The open points show the PYTHIA predictions for the pure NSD
sample at the UA5 and NA22 energies.}
\end{figure}

\par
 The situation is much more
involved for the moments. In Fig.4 we show the second scaled moment
$c_2$ for two choices of the rapidity range: $\mid \eta \mid <0.5$
and $\mid \eta \mid <2.4$ ($2.5$ for lower energies). Apart from the
data and MC results for two tunings (which coincide practically for
lower energies) we show here the MC predictions for "true NSD"
events. As shown in Fig.4, the MC predictions for the NA22 data
(defined by their trigger) are quite far from the MC results for
"true NSD" events. This casts some doubts on the claims of selecting
the "true NSD" events from the data by triggers and corrections. For
UA5 data the difference is smaller. Note that the difference between
the PYTHIA predictions for UA5 and CMS at the same energy of $0.9$
TeV results from the different definition of NSD events.
\par
The agreement of the model with data is unsatisfactory: the moments
are underestimated at NA22 and UA5 energies for both $\eta$ ranges.
However, the differences are not too big and the energy dependence
is qualitatively correct. The most important feature of the results
is the sudden increase of differences between the results from two
tunings at highest energies. This suggests that the reliable
measurement of the multiplicity distribution at $7$ TeV should fix
the tuning well enough to allow for a significant test of the model
from other data.

\begin{figure}[h!]
\centerline{ \epsfig{figure=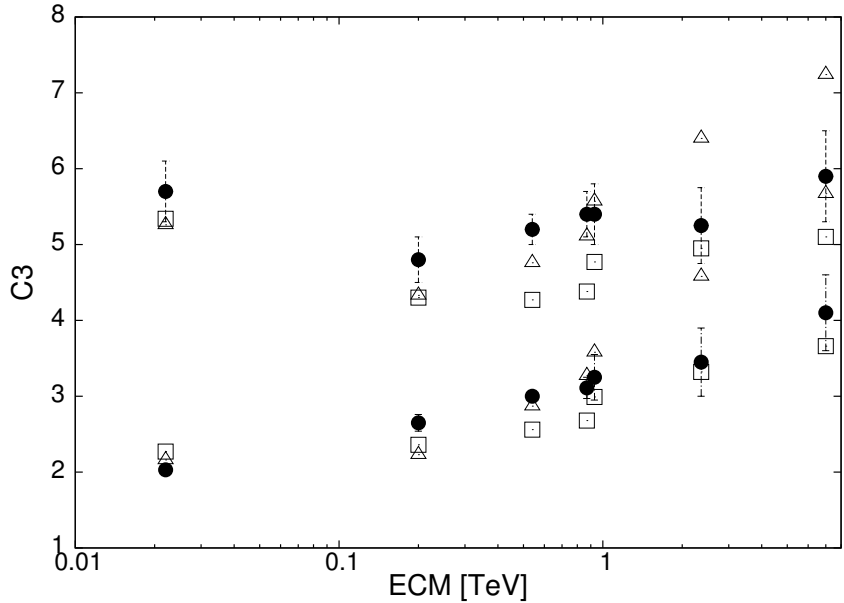,height=8.0cm}}
\caption{\footnotesize \label{C3} The $c_3$ moment in the $\mid \eta
\mid < 0.5$ and the $\mid \eta \mid < 2.4$ range from the NA22, UA5
and CMS data (full circles with error bars). PYTHIA 8.142
predictions with default and 8.135 tunings are shown as open squares
and triangles, respectively. For transparency, the data points for
UA5 and CMS and the PYTHIA predictions at $0.9$ TeV are slightly
shifted to lower and higher energy, respectively.}
\end{figure}

\begin{figure}[h!]
\centerline{ \epsfig{figure=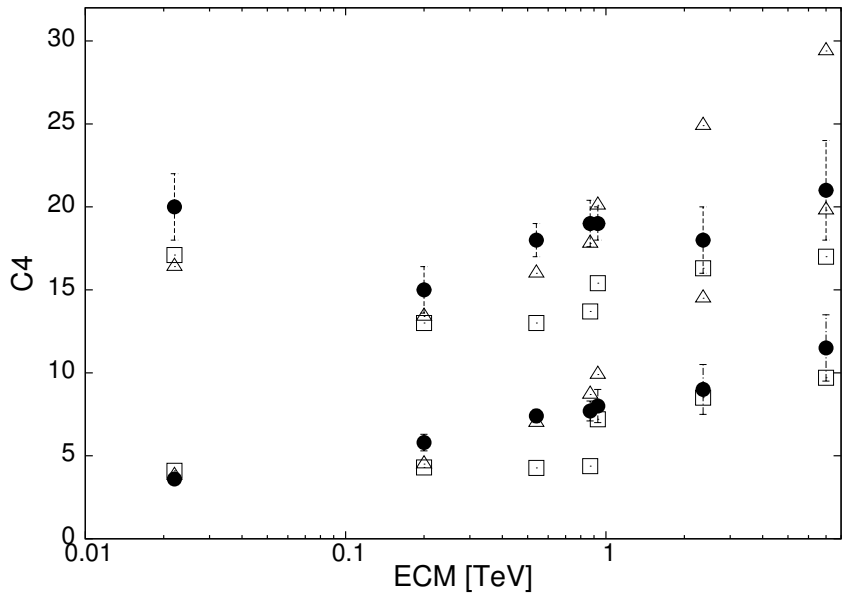,height=8.0cm}}
\caption{\footnotesize \label{C4} The $c_4$ moment in the $\mid \eta
\mid < 0.5$ and the $\mid \eta \mid < 2.4$ range from the NA22, UA5
and CMS data (full circles with error bars). PYTHIA 8.142
predictions with default and 8.135 tunings are shown as open squares
and triangles, respectively. For transparency, the data points for
UA5 and CMS and the PYTHIA predictions at $0.9$ TeV are slightly
shifted to lower and higher energy, respectively.}
\end{figure}

\par
In Fig.5 we show the third scaled moment for the rapidity ranges
$\mid \eta \mid <0.5$ and $\mid \eta \mid <2.4$ ($2.5$ for lower
energies). We see that the pattern is similar to that of the second
moment. For the NA22 data the value of $c_3$ is again much higher
than from the smooth extrapolation of the higher energy data,
suggesting that the trigger does not select well the NSD events. The
agreement of PYTHIA with low energy data is not quite satisfactory,
but the energy dependence is qualitatively correct. The LHC data are
bracketed by two versions of PYTHIA 8.142 tunings, which differ
strongly at highest energies.
\par
In Fig.6 the fourth scaled moment is shown for the rapidity ranges
$\mid \eta \mid <0.5$ and $\mid \eta \mid <2.4$ ($2.5$ for lower
energies), respectively. Again, the pattern is similar. Note that
the relative experimental uncertainties are almost the same. This
makes the discrepancies at lower energies less significant.
\par
In general, the qualitative agreement of PYTHIA with the energy
dependence of the mutiplicity distributions should be regarded as
acceptable.

\section {Conclusions and outlook}
We have extended our former analysis of the multiplicity
distributions at the LHC energies \cite{KFRW} using the new version
of MC generator (PYTHIA 8.142 with two tunings) and calculated the
scaled factorial cumulants from the scaled moments and average
multiplicities for the model and ALICE data. We have also
investigated the energy dependence of the average multiplicity and
the three lowest scaled moments in the wide energy range, comparing
the PYTHIA results with the NA22, UA5 and CMS data.
\par
We have found that the fast increase of the central density of
charged hadrons at LHC energies agrees quite well with the model
predictions. The use of factorial cumulants should facilitate the
fixing of the tuning parameters. The energy dependence of the scaled
moments is qualitatively well described in the wide range covering
the SPS, CERN collider and LHC energies.
\par
There is a large difference between the PYTHIA results with two
tunings at the highest energies. This suggests that the multiplicity
distributions from LHC are well suited to fix the tuning of MC
generators. Other data could be then compared with the model
predictions.

\end{document}